\documentclass[a4paper,11pt]{article}

\usepackage{pos}
\newcommand{\el}{\textit{e}$^-$ }

\title{Calibration of the Upgraded ALICE Inner Tracking System}

\manuallySeparateAuthors
\author*[a]{Andrea Sofia Triolo}
\author{, for the ALICE Collaboration}

\affiliation[a]{CERN, \\ Geneva, Switzerland}

\emailAdd{andrea.sofia.triolo@cern.ch}

\abstract{The ALICE Experiment has replaced its Inner Tracking System with a 7-layer pixel-only tracker made out of more than 24000 monolithic active pixel sensor chips, in order to fulfill the requirements of the physics program of the LHC Run 3. The upgraded Inner Tracking System (ITS2) has been installed in the ALICE experiment during the LHC long shutdown 2 and has started to take data with the beginning of Run 3 in July 2022, with proton-proton collisions at $\sqrt{s}$ = 13.6 TeV. With its 12.5 billion pixels it is the largest pixel detector installed in a high energy physics experiment to date. To guarantee stable operation and a consistently high data quality, a regular calibration of the detector has to be performed. The main part of the calibration program consists of a tuning and subsequent measurement of the pixel thresholds and a determination of the noisy channels. In particular the complexity of the threshold scan depends linearly on the number of pixels, which is why the threshold scan of the ITS2 is an unprecedented challenge. This work describes the architecture of the calibration framework, which has been developed using the detector control system of the ITS2 and the ALICE data processing layer. Results of first threshold and noise calibrations done in situ are shown as well.}

\FullConference{
  12-16 December 2022\\
  Santa Fe, New Mexico, USA}

\begin{document}
\maketitle

\section{The Upgraded ALICE Inner Tracking System}

\begin{figure}[h!]
    \centering
    \includegraphics[width=15cm]{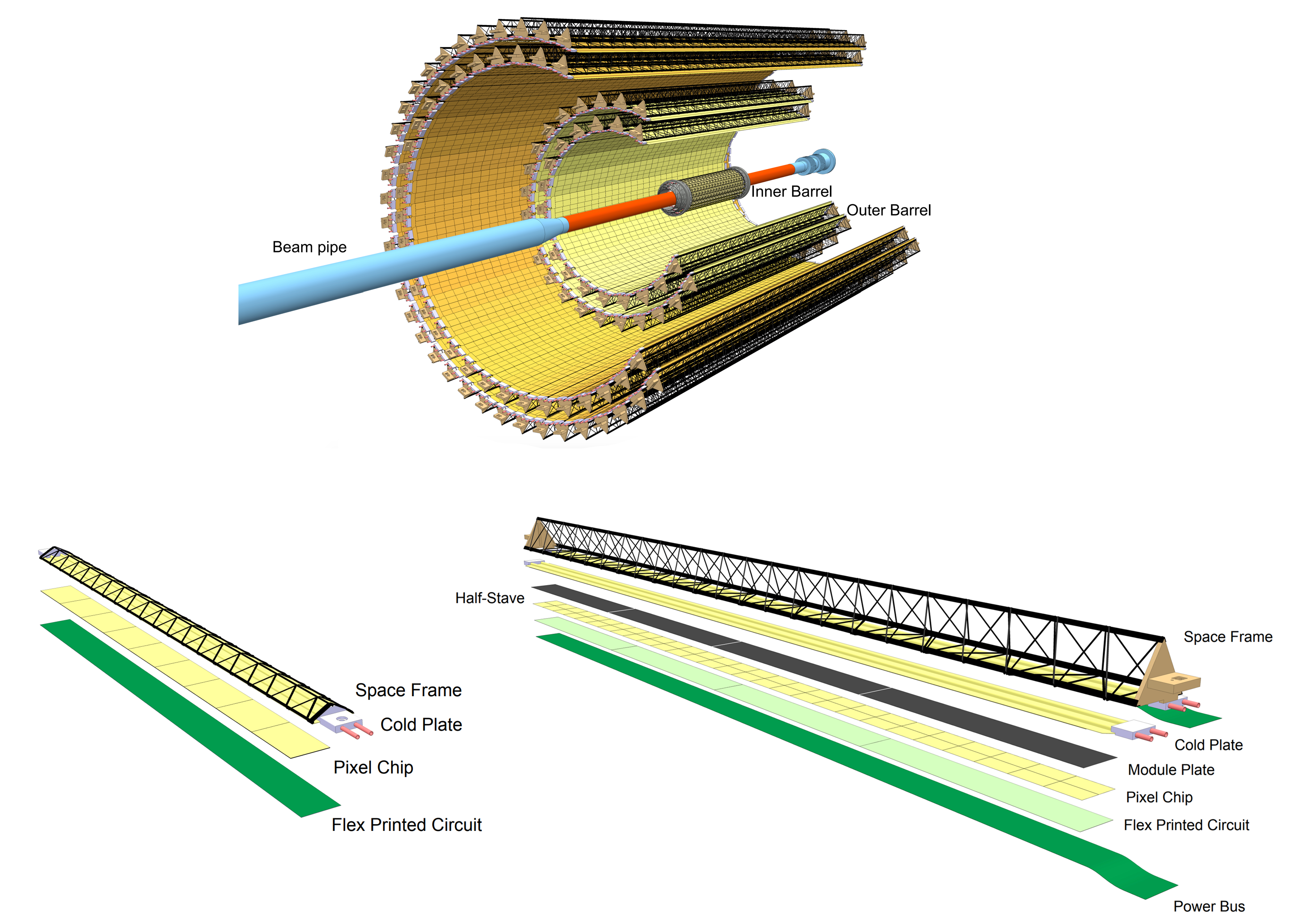}
    \caption{Schematic view of the ITS2 (top) and the constituent staves (bottom): Inner Barrel stave on the left, Outer Barrel stave on the right \cite{TDR}.}
    \label{fig:its2}
\end{figure}

ALICE (A Large Ion Collider Experiment) is one of the four largest experiments at the Large Hadron Collider (LHC) at CERN, and it is dedicated to the study of strongly interacting matter.
\\ ALICE completed a major upgrade in 2022, aimed to fulfill the requirements of the Run 3 physics program of the LHC \cite{TDR}.
The main physics topics addressed and discussed in the ALICE Upgrade Letter of Intent \cite{LoI} require the measurement of particles at low transverse momenta, together with novel measurements of jets and their constituents. Moreover, many of the measurements in Pb-Pb collisions are characterized by a very small signal-over-background ratio, and they require also a significant improvement in vertexing and tracking efficiency at low transverse momentum. The upgrade will provide an increase of statistics of about two orders of magnitude with respect to the period previous to the Long Shutdown 2.
\\
One of the several features of the upgrade was the replacement of the beam pipe with a smaller one. Thanks to this, the detectors were brought closer to the interaction point.
Consequently, the innermost detectors themselves were modified to suit the new configuration. In particular, the Inner Tracking System was entirely changed during the upgrade.
\\ A schematic view of the ITS2 is shown in Figure \ref{fig:its2}. It is the innermost detector of ALICE, and its main purposes are the localization of the primary and secondary vertices and the reconstruction of particle tracks.

The ITS2 consists of 7 cylindrical layers of Monolithic Active Pixel Sensors (MAPS) named ALPIDE (ALice Pixel Detector), with radial coverage from 2.2 cm to 39 cm.
The 7 layers are divided into Inner Barrel (IB, layers 0-2) and Outer Barrel (OB, layers 3-6), which is subsequently divided in Middle Layers (ML, layers 3-4) and Outer Layers (OL, layers 5-6). Each layer is azimuthally segmented into units named staves, for a total of 192 staves. 
Figure \ref{fig:its2} bottom shows two schematics of an IB stave and an OB (ML) stave.
The IB staves consist only of 9 chips in a row, instead the OB staves are composed of 112 chips for the ML or 196 chips for the OL, grouped in 4 rows of 7 chips each called modules, then the ML staves are composed by 4 modules, the OL staves by 7 modules.
\\With a total of 24120 chips, and an area of 10 m$^2$ covered by the detector, the ITS2 is the largest MAPS-based detector in High-Energy Physics to date.
\\With this new setup, another important feature achieved was the reduction of the material budget of the ITS: now the values of the material budget are 0.36\% $X_0$/layer for the IB and 1.1\% $X_0$/layer for the OB. This will allow the tracking performance and momentum resolution to be significantly improved, especially for the IB layers, where resolution may be more likely affected by multiple Coulomb scattering in the material.

\section{ALPIDE chips}
\begin{figure}[h!]
    \centering
    \includegraphics[width=12cm]{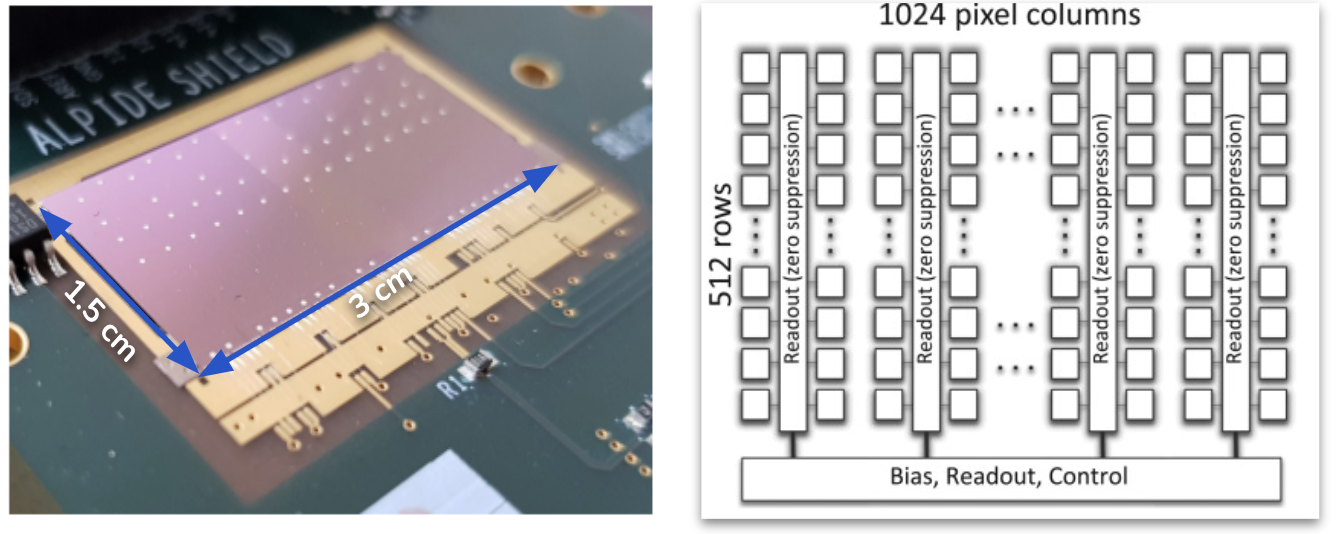}
    \caption{On the left: Photo of an ALPIDE chip. On the right: General architecture of the ALPIDE chip.}
    \label{fig:alpide}
\end{figure}

The ALPIDE chip (Figure \ref{fig:alpide}) is a CMOS MAPS developed on purpose for the ALICE Upgrade \cite{ALPIDE}. It is implemented in a 180 nm CMOS imaging process and fabricated on substrates with a high-resistivity (1-6 k$\Omega$cm) epitaxial layer. Its total power consumption is 40 mW/cm$^2$, and the input capacitance of the front-end is below 2 fF.
\\ The signal-sensing elements are n-well diodes with a diameter of approximately 2 $\upmu$m and an area typically 100 times smaller than the pixel cell area. The particular manufacturing process also provides a deep p-well layer that is exploited to implement the full CMOS circuitry in the active sensor area without compromising the sensing diodes charge collection.

The ALPIDE chips measure 15 $\times$ 30 mm$^2$ and include a matrix of 512$\times$1024 pixel cells, each one measuring 29.24 $\times$ 26.88 $\upmu$m$^2$.
Analog biasing, control, readout, and interfacing functionalities are implemented in a peripheral region of 1.2 $\times$ 30 mm$^2$. The periphery of the chip contains fourteen 8-bit analog DACs for the biasing of the pixel front-ends. 
Each pixel cell contains a sensing diode, a front-end amplifier and shaping stage, a discriminator, and a digital section. A hit is registered on a pixel only if the current signal induced in the input transistor and amplified is greater than a configurable threshold. In every pixel there is a pulse injection capacitor for the injection of test charges in the input of the front-end.
\\The sensor fulfils the experimental requirements of detection efficiency above 99\% and a spatial resolution of 5 $\upmu$m.

\section{ITS2 Calibration procedure}
To guarantee stable operation and consistently high data quality, a regular calibration of the detector has to be performed. The ITS2 configuration and calibration are managed by the ITS2 Detector Control System (DCS). Data extracted from the ITS through the O$^2$ First Level Processor (FLP) are sent to the Event Processing Nodes (EPN), which consist of 250 servers, each with 2 AMD Rome 32 core CPUs, 512 GB RAM, 8 AMD MI50 GPUs with 32 GB memory each, as well as a network interface. Here data are stored, reconstructed and processed online. While the experiment is not taking data, the EPN cluster is used as an analysis facility, reading data from the disk buffer for additional processing and physics analysis (offline processing).
\\The data readout and subsequent processing, including calibration, are integrated in the ALICE O$^2$ \cite{o2}, the ALICE official computing framework for Run 3.
\\
\\
A general ITS Calibration procedure involves the tuning of the pixel thresholds and the masking of the noisy pixels. The calibration is needed to guarantee a stable operational point and efficiency of the detector. There are three main calibration procedures for the ITS: threshold tuning, threshold scan and noise calibration. From the ITS2 Detector Control System (DCS) it is possible to configure the chips and control the scan execution, while the data readout and subsequent processing are integrated into the ALICE O$^2$. Other scans are available for detector studies.
\\
Data collected from the ITS2 are reconstructed and processed on the EPNs. The calibration of the ITS2 is a challenging procedure, especially from the computational point of view: with a total of 24120 chips and more than 12.5 billion pixels, the number of channels to be analyzed is enormous. To give an example, the number of EPNs required to perform the calibration is 40, out of 250 EPNs available in the ALICE farm: a threshold scan of the full detector results in about 3 $\times$ 10$^{13}$ hits, corresponding to 100 TB of raw hit data. To collect this huge amount of data the time needed is about 1 hour, and for this reason a scan of the full detector is generally used as reference and not performed on daily basis. Instead, at each beam dump, a threshold scan using only $\sim$2\% of pixels distributed uniformly on each chip is performed.

\subsection{Threshold tuning}
\begin{figure}
    \centering
    \includegraphics[width=15cm]{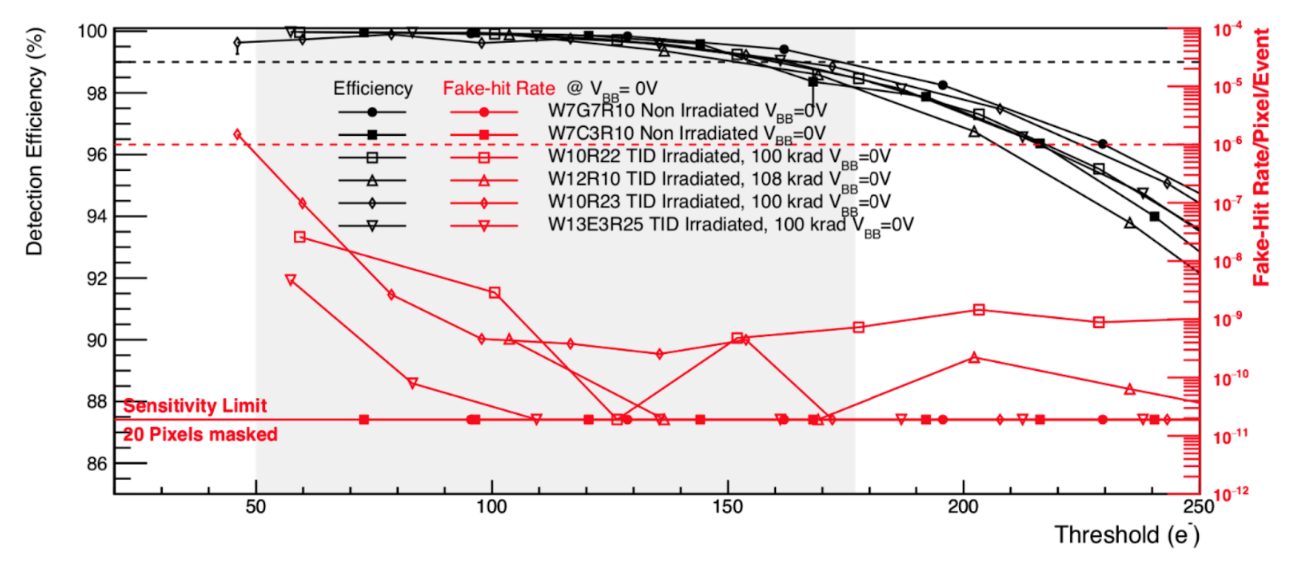}
    \caption{Experimental results obtained during beam tests. Detection efficiency and fake-hit rate are shown as a function of the chip threshold for different chips irradiation. }
    \label{fig:beamtest}
\end{figure}

The purpose of the threshold tuning is to set the operating point of the detector.
The threshold is influenced by the setting of two DACs, named VCASN and ITHR.
Setting the threshold of the detector means setting these two DACs. The target value chosen for the threshold setting is 100 \textit{e}$^-$, resulting from a compromise between having a good detection efficiency and a low fake-hit rate (FHR). As shown in Figure \ref{fig:beamtest}, the chip threshold should be maintained between 50 and 150 \el to keep the FHR below 10$^{-6}$ hits/events/pixel and the detection efficiency above 99\%, and thus 100 \el is taken as the optimal value at which the ALPIDE is fully efficient.

Figure \ref{fig:beamtest} also shows the importance of performing recurring threshold scans: if the threshold changes, a new tuning would be required to bring the threshold again at 100 \el.
\\
The tuning of the two DACs is performed by injecting an analogue pulse with the fixed amplitude of 100 \el and measuring the number of hits registered in the pixels. Then the DAC setting is changed and the injection of the fixed charge and the hit counting are repeated. This process is repeated 50 times for each DAC.
\\For each of the two DACs a separate scan is required. The first DAC to tune is VCASN, which provides a coarse tuning of the threshold. Then the ITHR tuning is performed to fine-adjust the threshold.  For each tuning, the pixel response is fitted with an error function to extract the 50\% point, which corresponds to DAC value to set to obtain a 100 \el threshold. 
\\The threshold tuning is performed at chip level, because the setting of VCASN and ITHR can only be made for the entire chip and not for the single pixels, so what is set is the average threshold for each chip. For this reason, only about 1\% of pixels for each chip are scanned for the tuning, for a total of $\sim$98 million pixels selected uniformly in the whole detector: only these pixels are enough to correctly find the VCASN and ITHR values for a correct threshold tuning.

\subsection{Threshold scan}
\begin{figure}
    \centering
    \includegraphics[width=15cm]{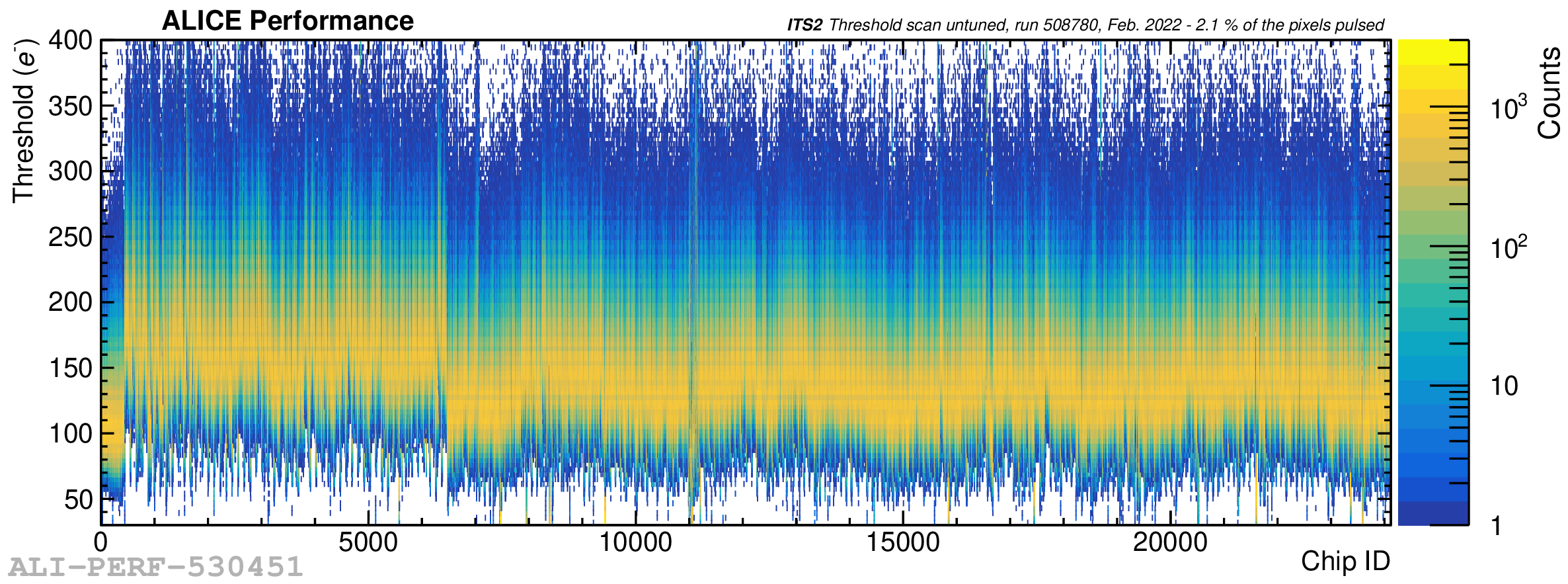}
    \includegraphics[width=15cm]{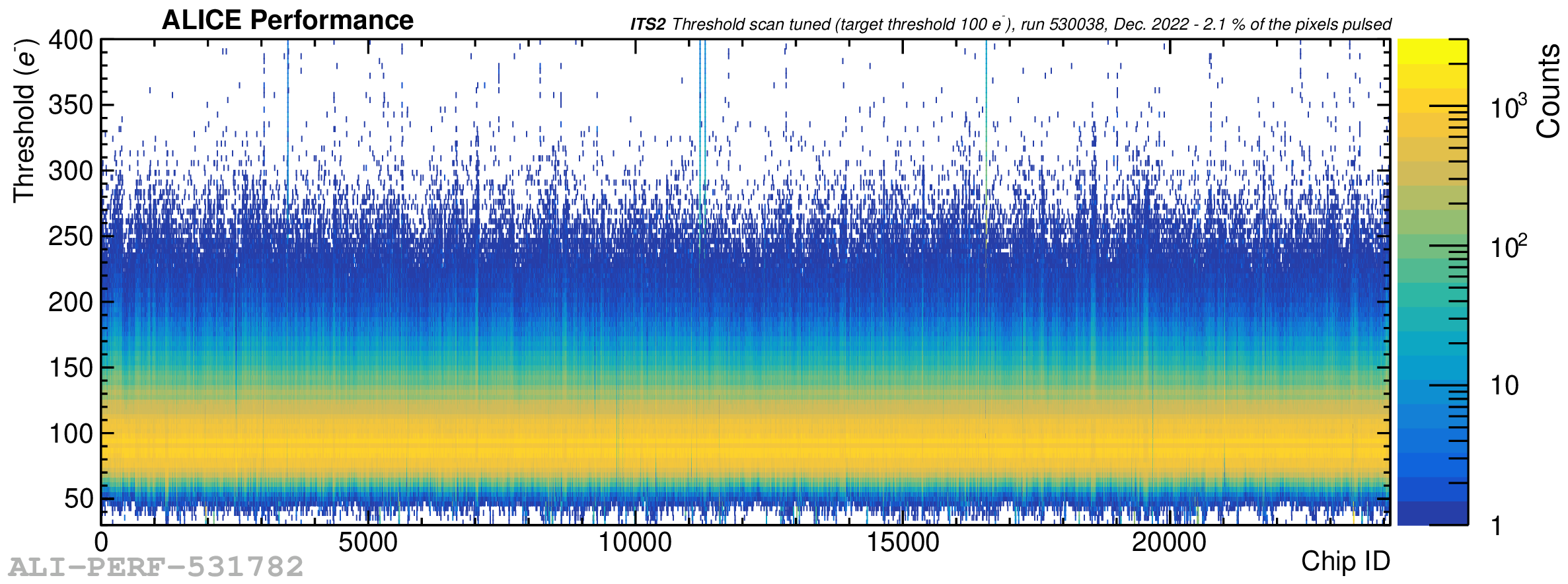}
    \caption{Top: Threshold distribution as a function of the chip ID for the full ITS2, before the tuning. Bottom: Threshold distribution  as a function of the chip ID for the full ITS2, after the threshold tuning to 100 \el.}
    \label{fig:thr}
\end{figure}

The threshold scan is performed to measure the threshold for each pixel of the detector. 
\\ This measurement is performed by injecting analog pulses ranging from 0 to 500 electrons into the pixels. In this case, the values of VCASN and ITHR found thanks to the threshold tuning are kept fixed for the chip: the purpose is indeed to test that the tuning is correct. The pixel response will result in an error function, that can be fitted to extract the 50\% point (threshold) and the $\sigma$ (ENC noise).
\\
Because of the huge amount of pixels constituting the detector, the threshold extraction for each pixel costs a lot in terms of time and resource consumption.
For this reason about 2\% of the pixels of the entire detector, taken uniformly from only 11 rows on each chip, are considered for this measurement, which means $\sim$271 million pixels for the whole detector; the scan with less pixels requires around 5 minutes to be completed, and the quality of the scan is not compromised from the reduced sample of pixels, since the chosen pixels are evenly distributed inside each chip and all the pixels behave almost the same way.
\\
In Figure \ref{fig:thr} it is possible to see the two distributions of the in-pixel threshold for each chip of the detector, before and after the tuning to 100 \textit{e}$^-$.
In the first case (top plot) it is clear that the threshold distribution is not uniform across all chips. In the bottom plot, obtained after the threshold tuning of the entire detector, it is evident that the average of every chip aligns with the target threshold of 100 \el at which the ALPIDE chips are known to be fully efficient, with an RMS of about 20 \el.

\begin{figure}
    \centering
    \includegraphics[width=11cm]{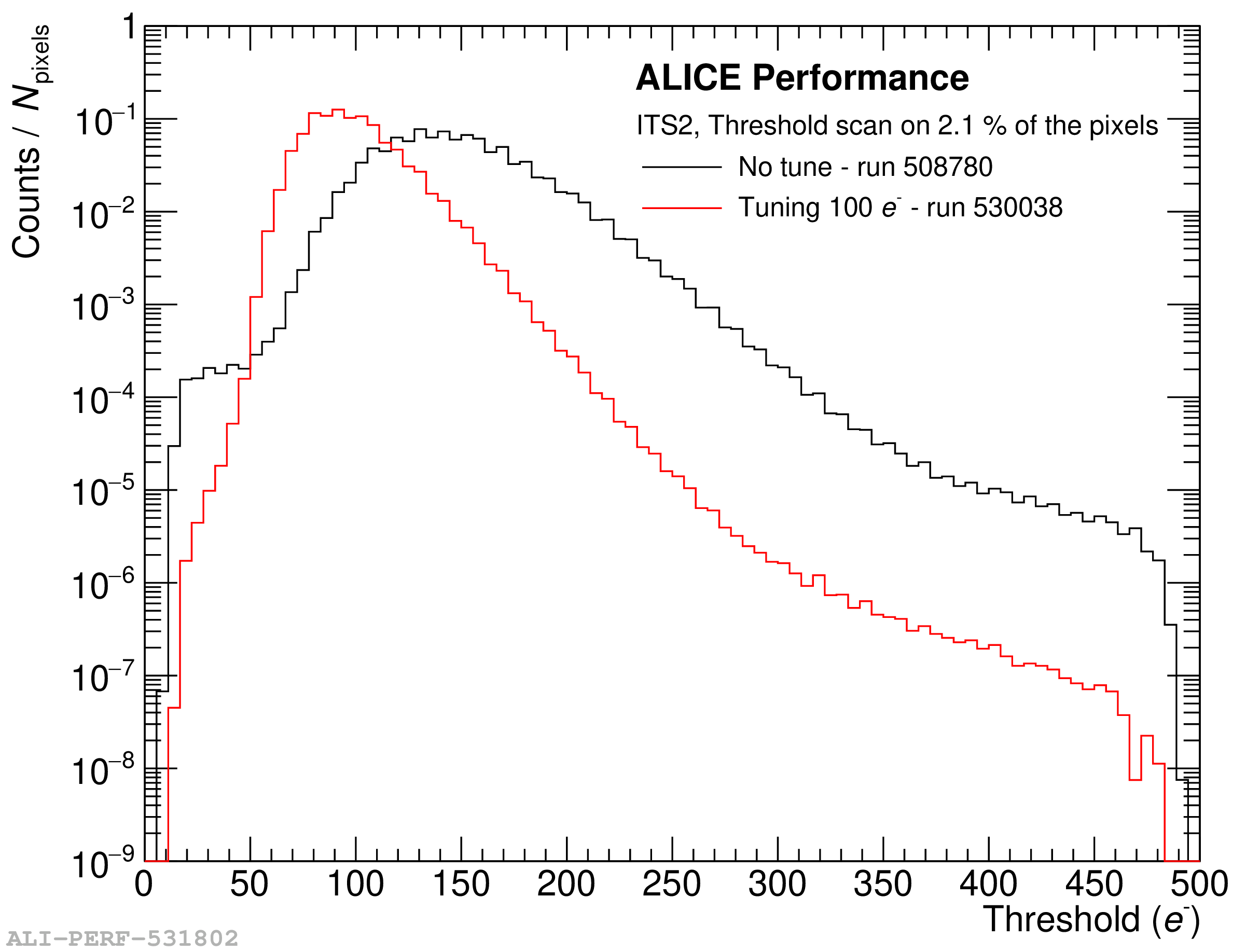}
    \caption{Threshold distribution with and without tuned threshold of ITS2. In the tuned case, the thresholds are tuned to 100 electrons.}
    \label{fig:1d_thr}
\end{figure}

The difference between the pixel thresholds before and after the tuning is directly visible in Figure \ref{fig:1d_thr}. In this plot, the black line refers to the not tuned case, and the red line refers to the tuned case, and both lines are normalized to the number of pixels considered in the scan. From the comparison of the two distributions it is clear that the tuning of the thresholds shifts the average point towards the target value (100 electrons) and reduces the number of pixels in the tail at large thresholds (>250 electrons) of about 2 orders of magnitude. A residual tail remains since we do not have control on every single pixel for the tuning, but only on the average per chip, which is our typical estimate of the chip threshold.

\begin{figure}
    \centering
    \includegraphics[width=15cm]{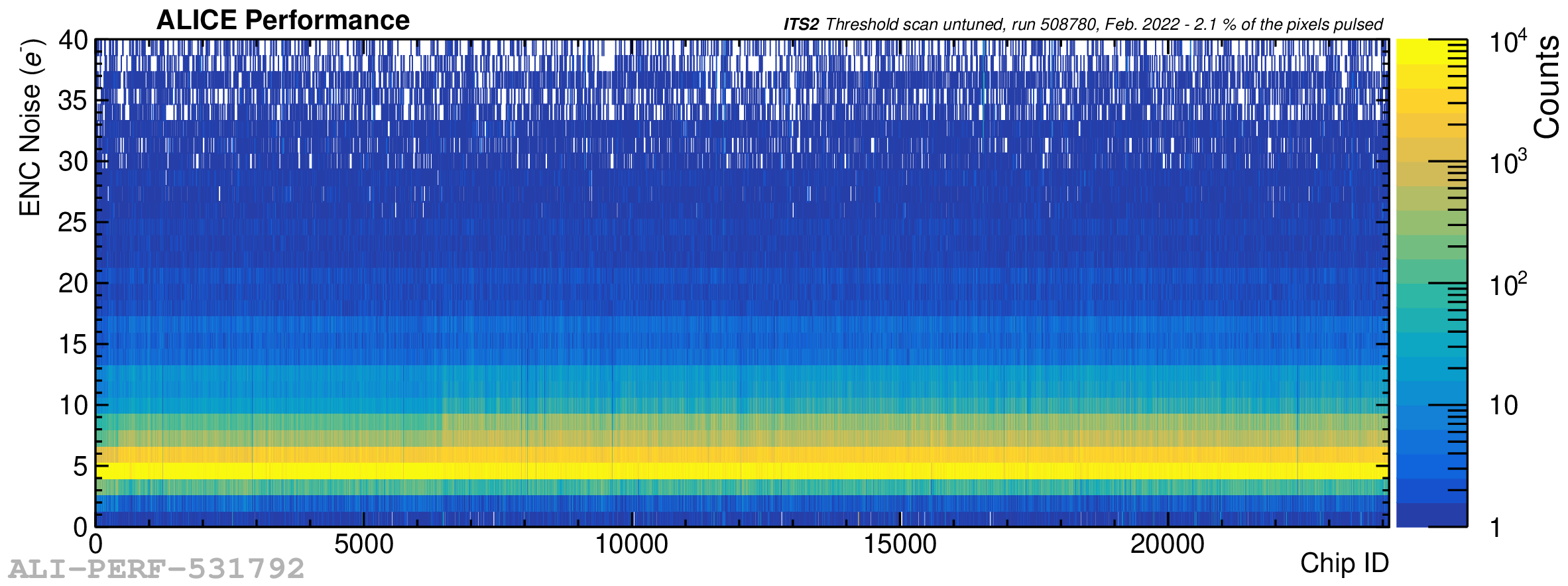}
    \caption{ENC Noise distribution as a function of the chip ID for the full ITS2.}
    \label{fig:noise}
\end{figure}

From the threshold scan it is also possible to measure the ENC noise.
In Figure \ref{fig:noise}, the typical ENC noise is of only 5 electrons and it is stable for all the chips, both before and after the tuning, so it is independent of the tuning of the detector.
\\Both threshold and noise measurements obtained are comparable with those obtained during the production phase.

\subsection{Noise calibration}
\begin{figure}
    \centering
    \includegraphics[width=15cm]{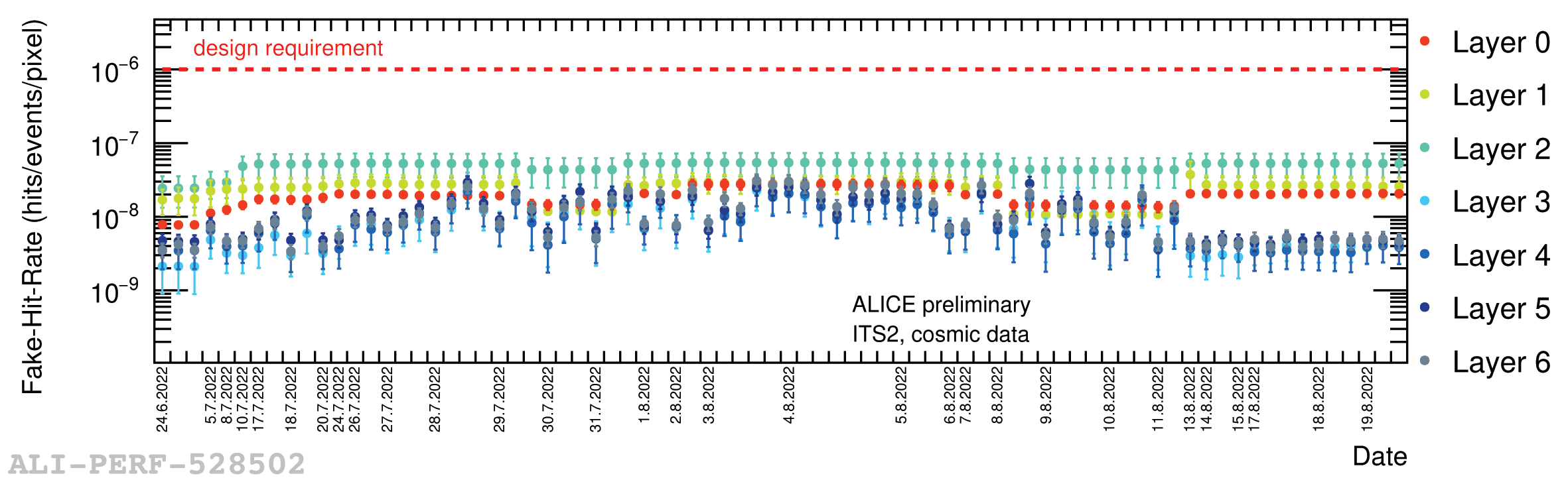}
    \caption{ITS2 average FHR trend as a function of date after masking 0.015\% of pixels of the entire detector. Each point shows a separate run from Run 3. The Average FHR was obtained by taking a mean value among each chip (HIC) FHR value for IB (OB) layer respectively. Errors are calculated as corresponding values of RMS/$\sqrt{N_{entries}}$}
    \label{fig:fhr}
\end{figure}

The purpose of the noise calibration is to tag and mask the noisy pixels of the detector, once the threshold is tuned.
\\ A noise calibration run consists in a standard data taking run without beam.
In order to mask the noisy pixels, a FHR threshold of 10$^{-2}$ hits/events/pixel for the IB and 10$^{-6}$ hits/events/pixel for the OB is set. As a result, all the pixels exceeding these values are tagged as noisy and masked to ensure a detection efficiency above 99\%. The difference between the two values of FHR chosen is driven by the choice to avoid losses of efficiency in the Inner Barrel. Moreover, considering the lower occupancy of the OB layers, the number of hits registered in the OB would be dominated by the noise if such a strict cut would not be applied.
\\ With these requirements, the number of pixels tagged as noisy and masked in the entire detector is around the 0.015\% out of 12.5 billion pixels (also including the pixels masked after the digital and analog scan that will be discussed in the next section), and they mostly come from the OB.
Figure \ref{fig:fhr} shows the average Fake-hit rate (FHR) for each layer after applying the mask. The overall value in average is stable over time and remains around 10$^{-8}$ hits/events/pixel, well below the 10$^{-6}$ hits/events/pixel imposed by design to achieve a good track reconstruction performance. 

\section{Monitoring scans: digital/analog scan, pulse-shape scan, DAC scan}

\begin{figure}[h!]
    \centering
    \includegraphics[width=15cm]{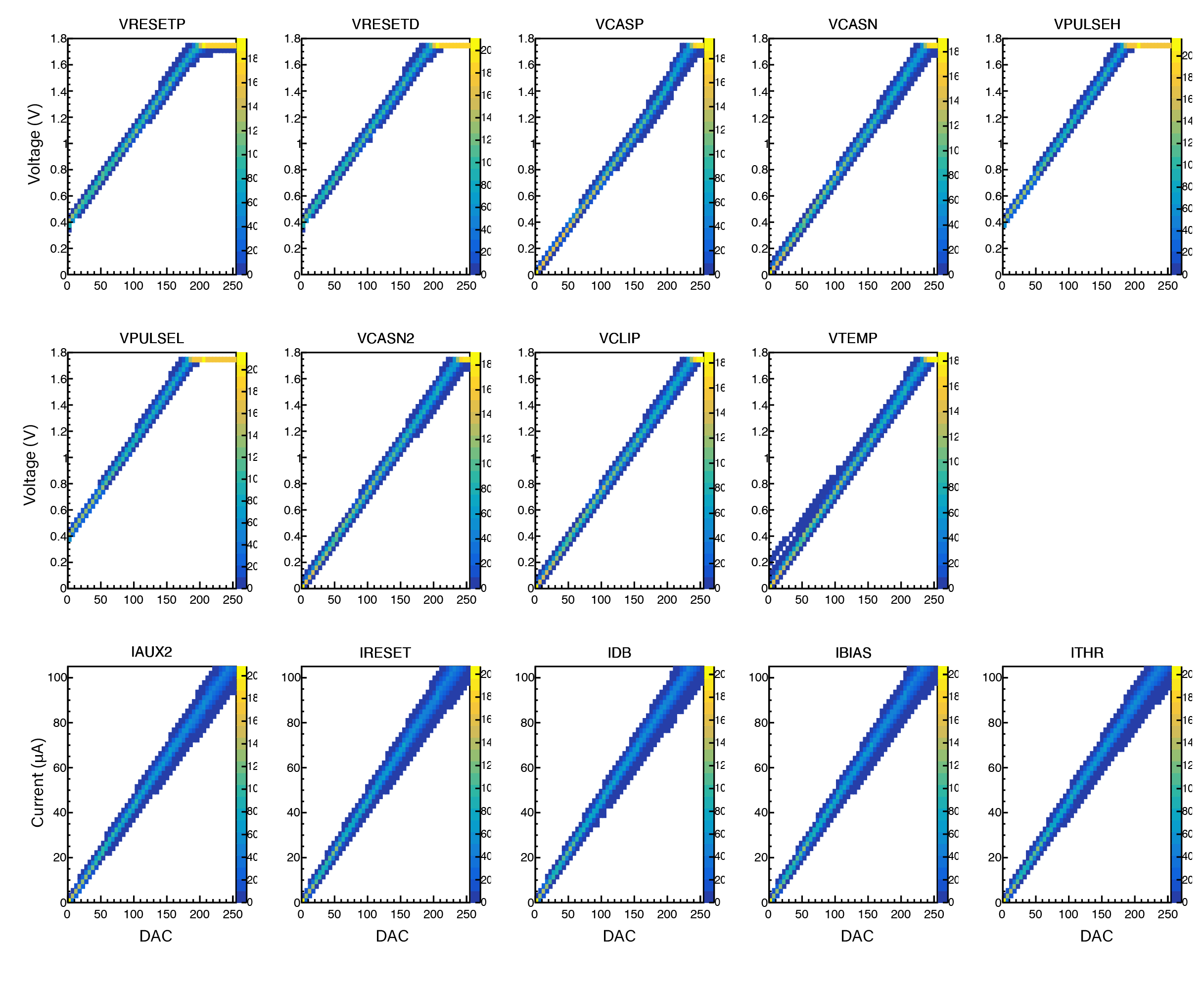}
    \caption{DAC scans for all the DACs for all the chips of the Inner Barrel. The analogue voltage and current values in output from the DAC are plotted as a function of the input digital values.}
    \label{fig:dacs}
\end{figure}

In addition to the main calibration scans discussed above, few other scans are available for monitoring and detector studies, for example some of them can be used to detect chip performance degradation due to radiation.
\\
\\
The Digital and Analog scans are used to mask the pixels with a problematic readout.
To spot the problematic pixels that have to be masked, the totality of pixels of the detector are pulsed 50 times (digitally or analogically, with an injected charge), and then a classification of the pixels is performed based on their response to the pulses: pixels with 0 hits registered are marked as "Dead", those with a number of registered hits between 0 and 49 are marked as "Inefficient", the ones with 50 hits registered are marked as "OK" and pixels with more than 50 hits registered are "Noisy". 
The pixels in the ALPIDE chips are organized in "double columns", two adjacent columns of pixels which share the same readout logic. If a double column contains more than 50 "Noisy" pixels, all its pixels are masked, in order to minimize the risk that the ALPIDE chip goes into busy status. The latter can happen because of a too large number of noisy pixels in a single double column, and also because it is assumed that the readout logic of that double column is not working properly.
\\ The amount of pixels masked this way is included in the 0.015\% of pixels masked for the entire detector.
\\
\\
The pulse-shape scan is used only for monitoring: it is possible to probe the shape of the pulse signal with a small strobe window to monitor the influence of this on the in-pixel threshold and to measure the time over threshold. The measurement is done sampling the pulse signal with different delays as a function of its amplitude and the delay of the trigger.
\\
\\
The DAC scans can be performed to monitor the on-chip DACs output. It is possible to perform a DAC scan for each of the 14 8-bit DACs present on the chip periphery. The scan is done by providing at the input of each DAC a digital value between 0 and 255 and reading the analog output. This scan could be performed for monitoring purposes, to verify that the linearity between the digital input and the analog output of the DACs is maintained.
Figure \ref{fig:dacs} shows a stable linearity of the DACs for all the chips of the Inner Barrel, meaning that the DACs were not modified by the radiation absorbed ($\sim$ 20 krad) in around 5 months of proton-proton collisions.

\section{Conclusions}

The ALICE Experiment has replaced its Inner Tracking System during the Long Shutdown 2, in order to fulfill the requirements of the Run 3 physics program of the LHC. The Upgraded detector is composed of 7 layers of monolithic active pixel sensor chips named ALPIDE. With
an area of 10 m$^2$ covered by the detector and 12.5 billion pixels, the ITS2 is the largest successfully operating MAPS-based detector in high-energy physics.
\\
The calibration of the detector consists of the tuning and measurement of the in-pixel threshold, and in spotting and masking the noisy pixels. The calibration procedure is very challenging because of the large number of channels that have to be calibrated: a threshold scan of the full detector results in about 3 $\times$ 10$^{13}$ hits, corresponding to 100 TB of raw hit data.
\\ From the start of the LHC Run 3 in July 2022 to the beginning of December 2022, ALICE accumulated a total absorbed dose of about 20 krad, a recorded luminosity of about 18 pb$^{-1}$ from proton-proton collisions, around 700 hours of data taking and a number of interaction recorder of the order of 10$^{12}$.
ITS2 threshold and noise remained stable during this period. The threshold is stable around 100 \el, and the fake-hit rate is stable around 10$^{-8}$ hits/events/pixel by masking only 0.015\% of the total number of pixels.
These stable results are comparable to single chip tests of threshold and noise performed during the various tests in the lab during the production phase.

\end{document}